%% file: main.tex
\documentclass[acmsmall]{acmart}

\AtBeginDocument{%
  }

\setcopyright{cc}
\setcctype{by-nc-nd}
\acmDOI{10.1145/3808131}
\acmYear{2026}
\acmJournal{PACMSE}
\acmVolume{3}
\acmNumber{FSE}
\acmArticle{FSE124}
\acmMonth{7}
\acmSubmissionID{fse26mainb-p365-p}
\received{2026-02-23}
\received[accepted]{2026-03-24}

\usepackage{amsmath,amsfonts}
\usepackage{algorithmic}
\usepackage{array}
\usepackage{adjustbox}
\usepackage[caption=false,font=normalsize,labelfont=sf,textfont=sf]{subfig}
\usepackage{stfloats}
\usepackage{url}
\usepackage{verbatim}
\usepackage{graphicx}
\usepackage{booktabs}
\usepackage{multirow}
\usepackage{colortbl}
\usepackage{microtype}

\usepackage{textcomp}
\usepackage{xcolor}
\usepackage[ruled]{algorithm2e}
\usepackage{makecell}
\usepackage{tcolorbox}
\usepackage{rotating}
\usepackage{longtable}
\usepackage{bbding}
\usepackage{tabularx}
\usepackage{pifont}
\usepackage{enumitem}
\usepackage{afterpage}

\input{tool}
\begin{document}

\title{RepoReasoner: Evaluating Repository-Level Code Reasoning Ability of Long-Context Language Models}
\author{Yanlin Wang}
\affiliation{
  \institution{Sun Yat-sen University}
  \city{Zhuhai}
  \country{China}
}
\email{yanlin-wang@outlook.com}

\author{Suiquan Wang}
\affiliation{
  \institution{Sun Yat-sen University}
  \city{Zhuhai}
  \country{China}
}
\email{wangsq37@mail2.sysu.edu.cn}

\author{Yanli Wang}
\affiliation{
  \institution{Sun Yat-sen University}
  \city{Zhuhai}
  \country{China}
}
\email{wangyli58@mail2.sysu.edu.cn}

\author{Bowen Zhang}
\affiliation{
  \institution{Sun Yat-sen University}
  \city{Zhuhai}
  \country{China}
}
\email{zhangbw53@mail2.sysu.edu.cn}

\author{Daya Guo}
\affiliation{
  \institution{Sun Yat-sen University}
  \city{Zhuhai}
  \country{China}
}
\email{guody5@mail2.sysu.edu.cn}

\author{Jiachi Chen}
\authornote{Corresponding author.}
\affiliation{
  \institution{Zhejiang University}
  \city{Hangzhou}
  \country{China}
}
\email{chenjiachi@zju.edu.cn}

\author{Zibin Zheng}
\affiliation{
  \institution{Sun Yat-sen University}
  \city{Zhuhai}
  \country{China}
}
\email{zhzibin@mail.sysu.edu.cn}

\newcommand{\ours}{Our Methods}
\begin{abstract}
Recent large language models (LLMs) have shown strong performance on software engineering tasks, yet most existing benchmarks evaluate code reasoning at the function level, where all relevant information is localized. This setting fails to reflect real-world development, which requires reasoning across multiple files and complex dependency structures. 
We introduce \bench, a benchmark for evaluating repository-level code reasoning. It assesses two complementary abilities: \textbf{Output Prediction}, which measures fine-grained, stateful execution reasoning across files, and \textbf{Call Chain Prediction}, which evaluates high-level architectural dependency understanding under noisy context. Our benchmark is constructed through a multi-stage pipeline that leverages dynamic tracing of \texttt{pytest} executions to obtain ground-truth call chains, along with LLM-based I/O rewriting to reduce memorization effects.
We evaluate seven state-of-the-art LLMs. Even under oracle context, the best-performing model achieves only 69.1\% Pass@1 on Output Prediction, indicating that cross-file reasoning remains a major challenge. In Call Chain Prediction, models exhibit high precision but low recall, suggesting limited multi-hop dependency understanding. Furthermore, performance drops on rewritten data reveal partial reliance on memorization, and longer contexts do not consistently improve results due to noise. 
These findings highlight fundamental limitations in current LLMs' repository-level reasoning and motivate future work on structured architectural understanding and cross-file inference.
\end{abstract}



\maketitle

\input{body.tex}

\begin{acks}
This work is supported by the National Natural Science Foundation of China (Grant No. 92582202, No. 62302534)
\end{acks}

\bibliographystyle{ACM-Reference-Format}
\bibliography{ref}

\end{document}

%% file: tool.tex
\newcommand{\boxmargin}{1.5mm}

\newtcolorbox{myboxc}{
    colback=yellow!10!white,    
    colframe=gray!50,           
    arc = 0pt, outer arc = 5pt,
    boxsep=5pt, left = 3pt, right = 0pt, top = 0pt, bottom = 0pt, 
    leftrule=3pt,               
    bottomrule=0pt, toprule=0pt, rightrule=0pt,
    left = \boxmargin, right = \boxmargin, top = \boxmargin, bottom = \boxmargin,
    before skip=5pt,   
    after skip=5pt    
}

\newcommand{\revision}[1]{\textcolor{black}{#1}}
\newcommand{\arxiv}[1]{\textcolor{black}{#1}}

\newcommand{\kmath}{$k$}

\newcommand{\bench}{RepoReasoner\xspace}

\usepackage{tikz}
\newcommand{\blackcircled}[1]{%
  \tikz[baseline=(char.base)]{%
    \node[shape=circle, fill=black, inner sep=1pt, text=white] (char) {#1};%
  }%
}

%% file: body.tex
\section{Introduction}
Recent advances in large language models (LLMs)~\cite{openai2025gpt41, guo2025deepseek, comanici2025gemini, hui2024qwen2.5, yang2025qwen3} have significantly improved their ability to handle complex software engineering tasks at the repository level~\cite{fan2023large, zan-etal-2023-large, zhang-etal-2023-repocoder, bairi2023codeplan, zhang2024codeagent, wang2024rlcoder}.
Consequently, there is growing interest in understanding whether LLMs can truly comprehend code structure and behavior across multiple files in repositories.  
To support this line of inquiry, several repository-level benchmarks~\cite{liu2024repoqa, hu2024coderepoqa, chen2025coreqa, li2025longcodeu} and code understanding benchmarks~\cite{liu2021codeqa, lee2022cs1qa, fu2023codeapex, manh2024codemmlu} have been developed to assess LLMs' reasoning abilities in cross-file dependency understanding.

Within this landscape, the evaluation of code reasoning ability has also evolved. Early benchmarks primarily focused on code generation~\cite{chen2021evaluating, austin2021program}, whereas more recent ones such as CRUXEval~\cite{gu2024cruxeval}, REval~\cite{chen2024reasoning}, and CodeSense~\cite{roy2025codesense} introduce tasks such as predicting program outputs and analyzing execution behavior. 
Despite these advances, a significant shortcoming persists: existing benchmarks operate almost exclusively at the \textbf{function level}. In these settings, all necessary information for solving a task is typically provided within a single, localized context. This approach fails to capture the complexity of real-world software development, where developers must continually reason about dependencies and logic scattered across the entire repositories. This creates a critical gap between current evaluations and real-world software engineering challenges.

\textbf{Benchmark Design.} To bridge this gap, we introduce \textbf{\bench}, a benchmark designed to evaluate repository-level code reasoning ability. Moving beyond self-contained code snippets, \bench assesses LLMs' ability to navigate, retrieve, and synthesize information distributed across multiple files, which is a skill essential for effective software engineering assistance. 

We propose two tasks: (1) \textit{Output Prediction}, which evaluates fine-grained, stateful reasoning by tracing complex, cross-file execution paths to predict a function's final output, and (2) \textit{Call Chain Prediction}, which assesses high-level architectural understanding by tasking it with identifying the correct sequence of files involved in an execution from a noisy context. \revision{These two tasks are designed to reflect two abilities required for repository-level reasoning: \textit{stateful execution reasoning} and \textit{architectural dependency reasoning}. Output Prediction evaluates whether models can propagate and combine information across multiple files while tracking program states to predict the concrete runtime result. Call Chain Prediction evaluates whether models can reconstruct the structural backbone of execution paths from noisy repository context. Notably, while Output Prediction may implicitly involve call-chain reasoning, Call Chain Prediction is not simply a subtask of Output Prediction: a model could predict the correct output without fully understanding the call chain, and conversely, it might correctly identify all invoked files but still fail to simulate the execution logic.}

\textbf{Evaluation.} We conduct an extensive evaluation of seven state-of-the-art LLMs, including open-source models such as DeepSeek-R1 and Qwen3, and closed-source models such as GPT-4.1-Mini and Gemini-2.5-Flash. According to the empirical results, we obtain the following key insights: \blackcircled{1} We find that even provided with perfect context, the best-performing model (DeepSeek-R1) achieves only 69.1\% Pass@1 score on Output Prediction, indicating that complex reasoning—not just retrieval—remains a fundamental challenge.
\blackcircled{2} 
We discover that LLMs have a limited capacity to understand a repository's high-level architecture, struggling to reconstruct the full call chains. Even top models struggle to accurately identify the invoked code file, exhibiting high precision but low recall, revealing weaknesses in multi-hop dependency tracing. 
\blackcircled{3} We show that models partially rely on memorization rather than reasoning. All models show a performance drop on our Input/Output rewritten (but logically identical) data, confirming their weaknesses when faced with unfamiliar code patterns.
\blackcircled{4} We find that simply increasing context length can be counterproductive. For some models (e.g., Qwen-2.5-Coder-14B), performance on Output Prediction decreases when expanding the context length from 10\kmath{} to 30\kmath{} tokens, as the harm from increased noise outweighs the benefit of potentially more complete information. 

Our main contributions are as follows:

\begin{itemize}
    \item We introduce a new benchmark \bench that shifts the paradigm of code reasoning evaluation from the function-level to the repository-level, better reflecting LLMs' reasoning capability in real-world code repositories.
    \item We present a robust benchmark construction pipeline that leverages dynamic analysis for ground-truth dependency tracing and I/O rewriting to prevent data contamination, ensuring a fair evaluation of a model's reasoning capabilities rather than its memorization.
    \item We conduct evaluations on a set of LLMs. Our findings reveal that cross-file reasoning is a primary bottleneck for repository-level reasoning and that current models may struggle with repository-level code reasoning, highlighting an area worth exploring for future research.
\end{itemize}

\section{Background \& Related Works}
\subsection{Long Context Language Models}
The evolution of LLMs is characterized by expanded context window capabilities, driving performance gains in long-context tasks like multi-document QA, long-form generation, and legal contract analysis. This progress also significantly enhances their abilities in a long code context, enabling more accurate and coherent processing of longer, intricate codebases, with ongoing advancements further extending their capacity for complex codebase processing.

For example, models like GPT-4.1~\cite{openai2025gpt41} and Qwen2.5-Coder~\cite{hui2024qwen2.5}demonstrated the early potential of this trend. Architectural innovations such as reinforcement learning to explicitly incentivize multi-step reasoning also underpin the strong long-context performance of models like DeepSeek-R1~\cite{guo2025deepseek}. However, the latest generation of models has dramatically accelerated this progress. Gemini 2.5~\cite{comanici2025gemini}, for instance, features a context window of up to 1 million tokens and employs a sparse mixture-of-experts (MoE) architecture to efficiently manage the vast context. Similarly, the recently released Qwen3~\cite{yang2025qwen3} series, particularly the flagship 235B model, trained on a massive dataset, showcases exceptional reasoning over long documents and complex instructions.

\subsection{Repository-Level Coding Research}
Repository-level code modeling has evolved from isolated function understanding toward holistic 
codebase comprehension, driven by the need to handle complex cross-file dependencies in modern 
software projects. 

Foundational works such as COCOMIC~\cite{ding-etal-2024-cocomic} and 
RepoCoder~\cite{zhang-etal-2023-repocoder} establish the importance of combining in-file and 
cross-file context through iterative retrieval frameworks. Building on these foundations, 
CodePlan~\cite{bairi2023codeplan} introduces systematic planning for multi-step repository tasks, 
while RepoFuse~\cite{liang2024repofuse}, RepoHyper~\cite{phan2024repohyper}, and 
RepoFormer~\cite{wu2024repoformer} advance context fusion and selective retrieval to improve 
cross-file information integration. More recent systems address training-level improvements: 
RLCoder~\cite{wang2024rlcoder} applies reinforcement learning to optimize repository-level 
completion\arxiv{, and AlignCoder~\cite{Jiang2025aligncoder} further improves retrieval alignment 
with target intent}. De-hallucinator~\cite{eghbali2024dehallucinator} mitigates hallucination through 
iterative grounding. Tool-integrated approaches such as CodeAgent~\cite{zhang2024codeagent} and 
context-aware frameworks~\cite{liao2023context,wang2024teaching} further extend these capabilities 
by incorporating external tools and autocompletion into the generation loop.
Retrieval-augmented generation forms another central pillar of repository reasoning. ReACC~\cite{lu2022reacc} 
and earlier neural retrieval methods~\cite{zhang2020retrieval,hayati2018retrieval} establish 
retrieval-enhanced generation paradigms, while syntax-aware and structure-aware 
strategies~\cite{zhang2023syntax,li2023acecoder} enable more precise, semantically grounded 
code retrieval. Documentation-centric methods like DocPrompting~\cite{zhou2022docprompting} tackle 
API-usage challenges by grounding generation in relevant documentation, and specialized techniques 
address private library integration~\cite{zan2022language}, bash code 
explanation~\cite{yu2022bashexplainer}, and edit-based summarization~\cite{li2021editsum}. 
Comprehensive surveys~\cite{parvez2021retrieval} systematize these retrieval-augmented paradigms, 
highlighting their central role in cross-file knowledge utilization\arxiv{, and there are also 
works studying what to retrieve for effective retrieval-augmented code 
generation~\cite{gu2025retrieve}}. 
\arxiv{At a larger scale, multi-agent systems~\cite{hong2024metagpt,huang2023agentcoder,qian2023communicative} demonstrate that coordinating specialized agents for planning, coding, and testing can substantially improve repository-scale code generation quality. Beyond completion and retrieval, the community has further proposed benchmarks targeting repository-level code translation~\cite{ou2025rustrepotrans,wang2024repotransbench} and secure code generation across real-world repositories~\cite{wang2026realsecbenchbenchmarkevaluatingsecure}.}

\subsection{Coding Ability Evaluation Benchmarks}

\arxiv{Previous works have proposed benchmarks to evaluate the coding ability of LLMs from various perspectives~\cite{zheng2025generalperformancedomain,gong2026cosqa+,zhang2025simpledevqabenchmarkinglargelanguage}.} To evaluate LLM code comprehension and reasoning capabilities in particular, numerous benchmarks have been proposed across different aspects of code intelligence. Earlier benchmarks~\cite{liu2021codeqa,lee2022cs1qa,fu2023codeapex} assess snippet-level understanding through question answering and educational programming tasks, \arxiv{while more advanced reasoning 
benchmarks~\cite{gu2024cruxeval,xu2025cruxevalx,chen2024reasoning,liu2024codemind,wei2025equibench,xie2025core} evaluate execution prediction, runtime behavior reasoning, semantic equivalence, and static dependency analysis. Scalable evaluation paradigms~\cite{zhang2024multiple,cobbe2021gsm8k,manh2024codemmlu,ma2024speceval,andryushchenko2024CodeQuestionAnswering,fu2024scratcheval} further broaden coverage through multiple-choice and specification-driven settings but largely remain at the function level.}

\revision{In addition to code understanding benchmarks, SWE-bench~\cite{jimenez2024swebench} 
represents an important line of work that evaluates LLMs' \textit{end-to-end} issue resolving 
ability. SWE-bench integrates multiple capabilities including repository navigation, code repair, 
patch validation, and test execution \arxiv{, and OmniGIRL~\cite{guo2025omnigirl} further extends this setting to multilingual and multimodal issue resolution.}} 
With the emergence of long-context models, evaluation has shifted toward repository-level settings, 
where RepoQA~\cite{liu2024repoqa}, CodeRepoQA~\cite{hu2024coderepoqa}, 
CoReQA~\cite{chen2025coreqa}, and LongCodeU~\cite{li2025longcodeu} require cross-file retrieval 
and reasoning over entire repositories. 
\revision{However, unlike prior benchmarks that either emphasize end-to-end issue resolution or 
primarily frame repository reasoning as retrieval and question answering, our benchmark explicitly 
isolates and evaluates two core capabilities: stateful cross-file execution reasoning and 
architectural dependency reconstruction under noisy repository context.}

\section{\bench}
This section details the construction of our benchmark, which is designed to address a critical gap in current code reasoning benchmarks: the assessment of reasoning capabilities in a long-context, repository-level setting.
We begin with a formal definition of our evaluation tasks, a step-by-step description of our data construction pipeline, and a comparative analysis that positions our benchmark within the landscape of existing code intelligence evaluations.

\subsection{Benchmark Tasks}
To assess repository-level code reasoning, we propose two tasks that address limitations in existing function-level evaluations. Our approach targets the dual reasoning processes required in software engineering: tracing specific execution paths (micro-level reasoning) and understanding architectural structure (macro-level reasoning). Both capabilities are fundamental to effective code comprehension. We introduce \textbf{Output Prediction} and \textbf{Call Chain Prediction} tasks to evaluate these two aspects independently.

\begin{figure}[ht!]
    \centering
    \includegraphics[width=\textwidth]{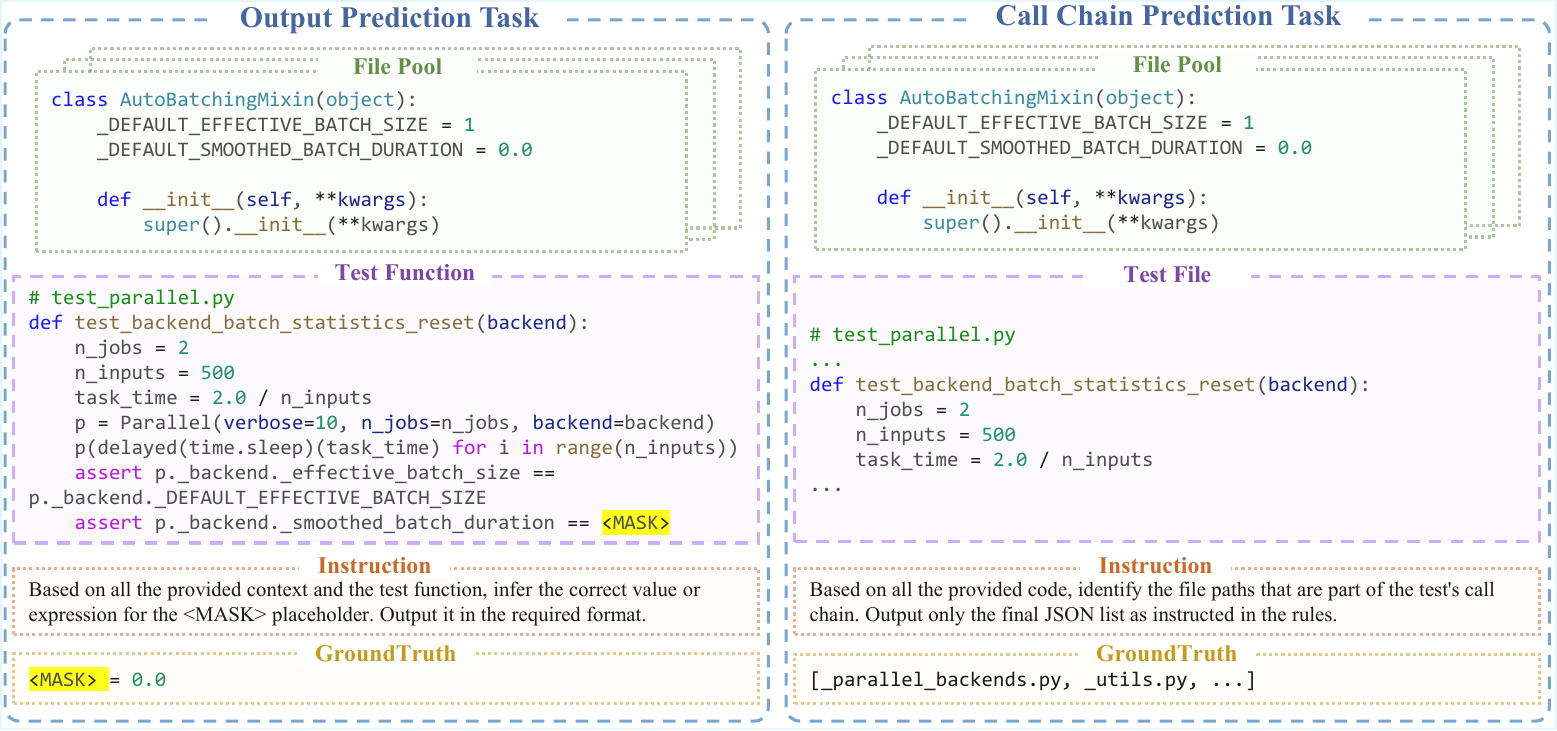}
    \caption{An illustration of the two tasks in \bench. The left panel shows an example of the \textbf{Output Prediction} task, where the model needs to predict the masked value with the test function and related files. The right panel depicts the \textbf{Call Chain Prediction} task, requiring the model to identify the correct files involved in the execution from a file pool.}
    \label{fig:task_examples}
\end{figure}

\subsubsection{Output Prediction}
This task is designed to evaluate a model's capacity for fine-grained, stateful code execution simulation, moving far beyond the scope of single-function analysis. In any non-trivial repository, the behavior of a function is rarely self-contained. Its final output is conditioned by a web of interconnected factors: class constructors setting initial object states in one file, configuration constants defined in another, and parent classes providing base methods from yet another part of the hierarchy. To succeed, a model must demonstrate a holistic understanding of the repository's dynamic state. This requires accurately resolving object-oriented features like inheritance and polymorphism to trace method calls correctly, tracking the lifecycle of objects from instantiation to use, and understanding how global-scope variables or external configuration files alter the execution flow.

This mirrors the cognitive process of a developer during critical tasks like debugging or root cause analysis. When faced with an unexpected output, a developer must mentally trace the program's execution, stepping through multiple files and holding variable states in mind to pinpoint the logical flaw. By forcing models to synthesize this distributed information to predict a precise outcome, we directly test their ability to perform this deep inferential reasoning, distinguishing them from models that rely on superficial pattern matching.

To operationalize this, we ground the task in the existing unit tests of a repository. As shown on the left side of Figure~\ref{fig:task_examples}, we select a test function (e.g., \verb|'test_function'|) containing a ground-truth assertion like \verb|'assert Expression == Output'|. We then transform this into a cloze-style question: \verb|'assert Expression == MASK'|. The model is provided with the information from repository as context and is tasked with predicting the content of the \verb|'MASK'|.

The core challenge lies in evaluating the \verb|'Expression'|, which is intentionally designed to trigger a chain of cross-file function calls and object interactions. The model must first parse the assertion, identify the entry point, and then navigate the code in the context to find definitions of all invoked functions, classes, and variables. Finally, it must simulate the execution path, respecting the logic and state changes along the way, to compute the final value. The target prediction for \verb|'MASK'| is flexible and can be a literal value (e.g., \texttt{42}, \texttt{`Completed'}), a symbolic identifier (e.g., a variable name like \verb|'expected_status'|), or another valid expression. This variety prevents models from succeeding by simply guessing common default values and ensures they have a true semantic understanding of the code's execution.

\subsubsection{Call Chain Prediction}
Related to the micro-level focus of Output Prediction, this task evaluates a model's macro-level architectural comprehension. Its goal is to assess whether a model can construct a mental map of a repository's structural dependencies, a crucial skill for navigating large and unfamiliar codebases. This directly simulates a foundational developer behavior: when trying to understand a feature or fix a bug, an engineer often uses IDE tools like ``Go to Definition'' and ``Find Usages'' to trace the flow of control and data. This process builds a cognitive model of the call stack and component interactions.

A strong performance in this task requires more than simple text-based searching for function names. The model must correctly parse import statements, resolve multi-hop dependencies across files, and handle complexities like dynamic dispatch, where the specific method called depends on an object's runtime type. This high-level architectural awareness is indispensable for many software engineering activities, such as performing impact analysis before a major refactoring, onboarding new team members by summarizing key code paths, or identifying critical components for performance optimization. By challenging the model to distinguish relevant files from a sea of irrelevant ones, we measure its ability to focus on the essential structure of the code.

The task setup, illustrated on the right side of Figure~\ref{fig:task_examples}, centers on a target test function from a specific file (e.g., \verb|test_file.py|). We then provide the model with a \verb|file_pool|—a curated collection of source code files from the same repository. This pool is intentionally designed to be noisy. It contains all the ``signal'' files that are part of the ground-truth call chain established by a dynamic execution trace of the test function. Crucially, it also includes a number of ``noise'' or ``distractor'' files, which are other valid but unrelated source files from the project.

The model's objective is to analyze this entire pool and predict the list of files that are actually invoked during the execution of the target test, in the form of \texttt{[f\_1.py, f\_2.py, ..., f\_n.py]}. It elevates the task from simple file classification to a more challenging test of control-flow understanding. The more correct predicted files in the list, the more likely it is that the model knows which files are related from the functionality and logic, rather than just navigating by the import statements. This directly measures the model's capacity to trace execution flow across file boundaries while robustly filtering out distracting, irrelevant information—a foundational skill for comprehending any complex software architecture.

\subsection{Benchmark Construction Pipeline}
\begin{figure}[t]
    \centering
    \includegraphics[width=\textwidth]{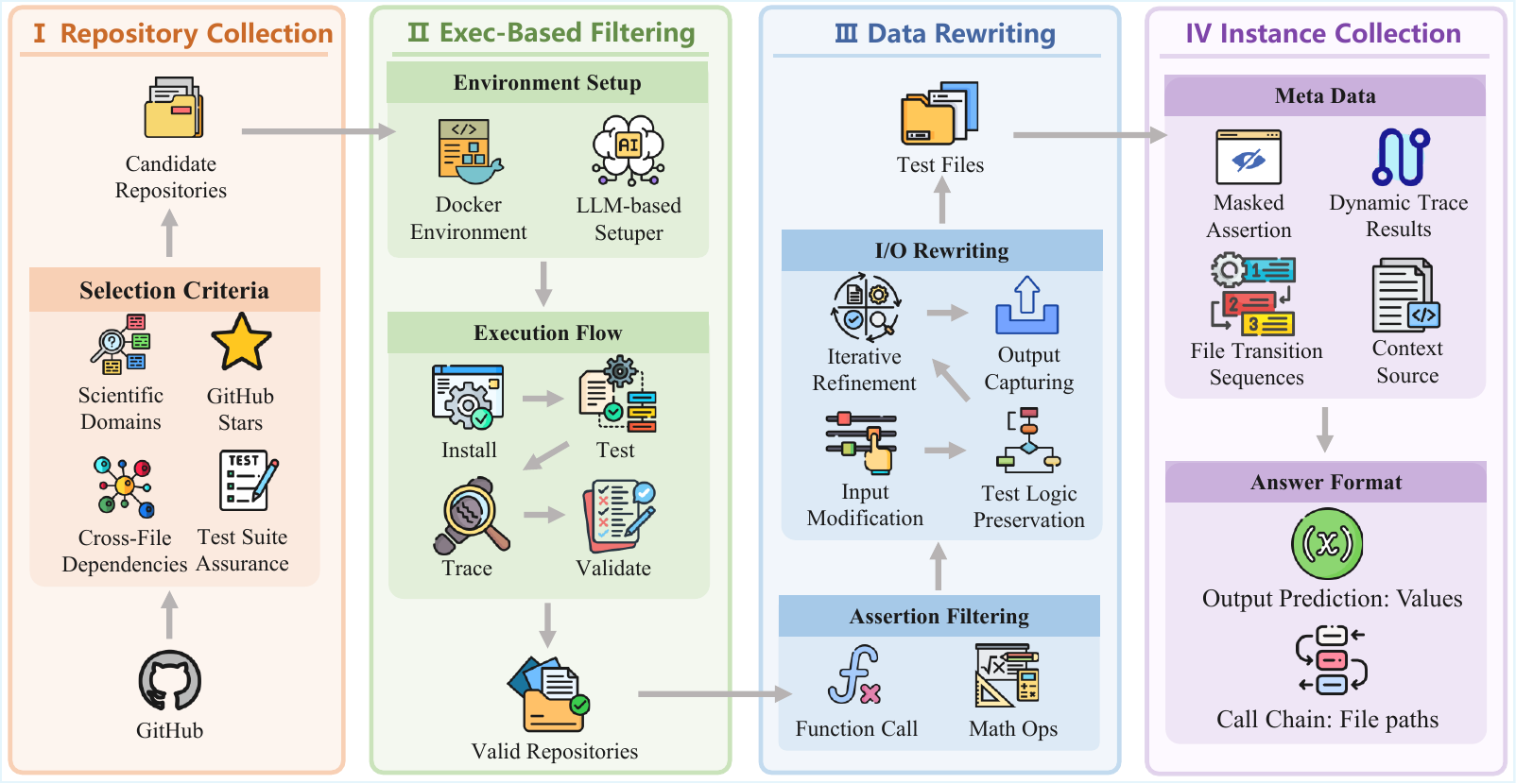}
    \caption{An overview of \bench construction pipeline.}
    \label{fig:pipeline}
\end{figure}

\noindent Our benchmark construction follows a four-stage pipeline that progressively filters and processes repositories to create evaluation instances. Figure~\ref{fig:pipeline} illustrates the complete workflow.

\subsubsection{Stage I: Repository Collection}
\noindent We initiate our process by curating a diverse collection of Python repositories from over 20 scientific domains including machine learning, astrophysics, and parallel computing. The selection criteria emphasize both code quality and structural complexity. Candidate repositories are evaluated based on community engagement metrics such as GitHub stars and active maintenance, alongside strong software engineering principles. Each repository must contain standard configuration files (\texttt{setup.py} or \texttt{pyproject.toml}) to facilitate automated and reproducible environment setup. We prioritize libraries for data analysis, numerical computing, and scientific applications to ensure that tasks involve complex data transformations and function calls from which concrete numerical values can be extracted as ground truth. A primary selection criterion is the exhibition of deep and complex cross-file dependency graphs, ensuring that tasks genuinely require repository-level context understanding. To guarantee code reliability, we require projects to have an active development history, a comprehensive test suite implemented with \texttt{pytest}, and a record of passing all tests.

\subsubsection{Stage II: Execution-Based Filtering}
\noindent To capture real cross-file interactions, we perform dynamic analysis within isolated and reproducible Docker~\cite{docker} environments. The Docker container is built upon an Ubuntu 22.04 image, equipped with a complete Python 3 toolchain and pre-installed dependencies like \texttt{pytest} and \texttt{hunter}~\cite{hunter} to ensure consistency. Within this controlled environment, a lightweight, LLM-powered tool driven by GPT-4 automates the setup and execution for each repository. To ensure robustness, the tool's interaction with the repository is constrained: a maximum of 15 interactions is allowed per repository to prevent infinite loops from configuration failures, and timeouts are set to 300 seconds for standard commands and 1800 seconds for the \texttt{pytest} execution to accommodate lengthy test suites. To trace the dependencies, our custom \texttt{pytest} plugin hooks into the execution engine, leveraging the \texttt{hunter} library for tracing Python code execution. This setup records every function call during the test runs, capturing its origin file, destination file, and the precise sequence of file-level transitions. This dynamic tracing yields a detailed call chain for each test function, including the number of file hops, which is superior to static analysis as it captures the precise, runtime-dependent control flow. This filtering reduces our dataset from 100 to 15 repositories that successfully pass all tests and generate traceable execution patterns.

\subsubsection{Stage III: Data Rewriting}
\noindent This stage is crucial for creating new evaluation instances that avoid models remembering ground truth during the training phase, with a primary focus on the Output Prediction task. The process ensures that models are evaluated on their reasoning ability rather than their capacity for memorization. We first develop a filtering mechanism to identify high-quality assertion statements suitable for task generation. This process combines regular expression matching, Abstract Syntax Tree (AST) parsing, and expression complexity analysis. We traverse the AST of each test function to extract all assertion nodes, then filter out low-quality candidates such as simple literal comparisons or direct variable-to-value checks (\texttt{assert x == y} or \texttt{assert x == 5}). We select for high-quality structures that involve function calls, attribute access, or complex mathematical operations, ensuring the assertion's value is the result of a non-trivial computation. To generate new instances unlikely to have appeared in any model's training data, we employ the same LLM-powered tool within the Docker environment. The tool follows an iterative refinement process: it first analyzes the original test function and designs a more complex test case by modifying the input variables, considering edge cases and different data types. The tool alters only the input values and the corresponding expected value in the assertion, leaving all other logic intact. The modified test is then re-executed. If an \texttt{AssertionError} occurs, the error message, along with the generated code, is fed back to the tool. The tool then fixes only the expected value in the assertion to match the actual output from the error trace, without changing the input again. This loop continues until the new test case passes. Another critical step is supplementing the ground truth: the original assertion might compare an expression to an identifier, but a model might correctly predict the literal value of that identifier. To ensure robust evaluation, we record both the identifier and its actual runtime value as acceptable answers.

\subsubsection{Stage IV: Instance Collection}
\noindent The final stage involves collecting all processed information and packaging it into a standardized, self-contained format for each benchmark instance. A single data point is structured to provide a complete evaluation scenario. Each finalized instance includes task specification with the full path to the target test file (\texttt{tests/modules/test\_core.py}), the name of the specific test function (e.g., \texttt{test\_complex\_logic}), and the exact code containing the masked assertion statement (e.g., \texttt{assert Expression == MASK}). For the Output Prediction task, we provide a list of all valid answers captured during Stage III, including both literal values and any equivalent identifiers or expressions (e.g., \texttt{'EXPECTED\_VALUE'} or \texttt{1.0}). For the Call Chain Prediction task, we include the ground-truth list of file paths that constitute the correct call chain. The ground-truth call chain is captured by our custom \texttt{pytest} plugin during the dynamic analysis in Stage II and serves as the source of context in our experiments and enables detailed, post-hoc case analysis of model behaviors and failures. Each instance is tagged with relevant metadata for traceability and analysis, including the source repository's name, the unique ID of the instance, and other relevant information such as the original, unmasked assertion. The final dataset contains 525 test functions across 169 test files from 14 repositories.
Based on the construction of the pipeline as detailed above, we curated a dataset comprising 858 samples for output prediction and 169 samples for call chain prediction.
\subsection{Characteristics of \bench}
Our benchmark introduces a paradigm shift in code reasoning evaluation by moving from the function-level to the repository-level. The fundamental differentiator between our work and benchmarks like CRUXEval or CodeSense is the scope of context. Those benchmarks operate at the function-level, providing all necessary information within a single, localized prompt to test a model's ability to reason over a given block of code. In contrast, our benchmark operates at the repository-level. The model is given many cross-file information and must independently search the most relevant information needed to solve the task, thus evaluating the more complex and practical skill of retrieving and synthesizing distributed information.

Similarly, while REval focuses on deep, intra-procedural reasoning by asking models to predict program states at a statement-by-statement granularity, our benchmark tests a different dimension. REval assesses a model's ability to be a meticulous ``code interpreter''. Our benchmark, on the other hand, evaluates its ability to be a ``code navigator'' and ``architectural analyst'', understanding the broader structure and integrating information from disparate parts of the codebase. The skills are related but distinct, with our work addressing a larger-scale of program comprehension.

\begin{figure}[ht!]
\centering
\includegraphics[width=\textwidth]{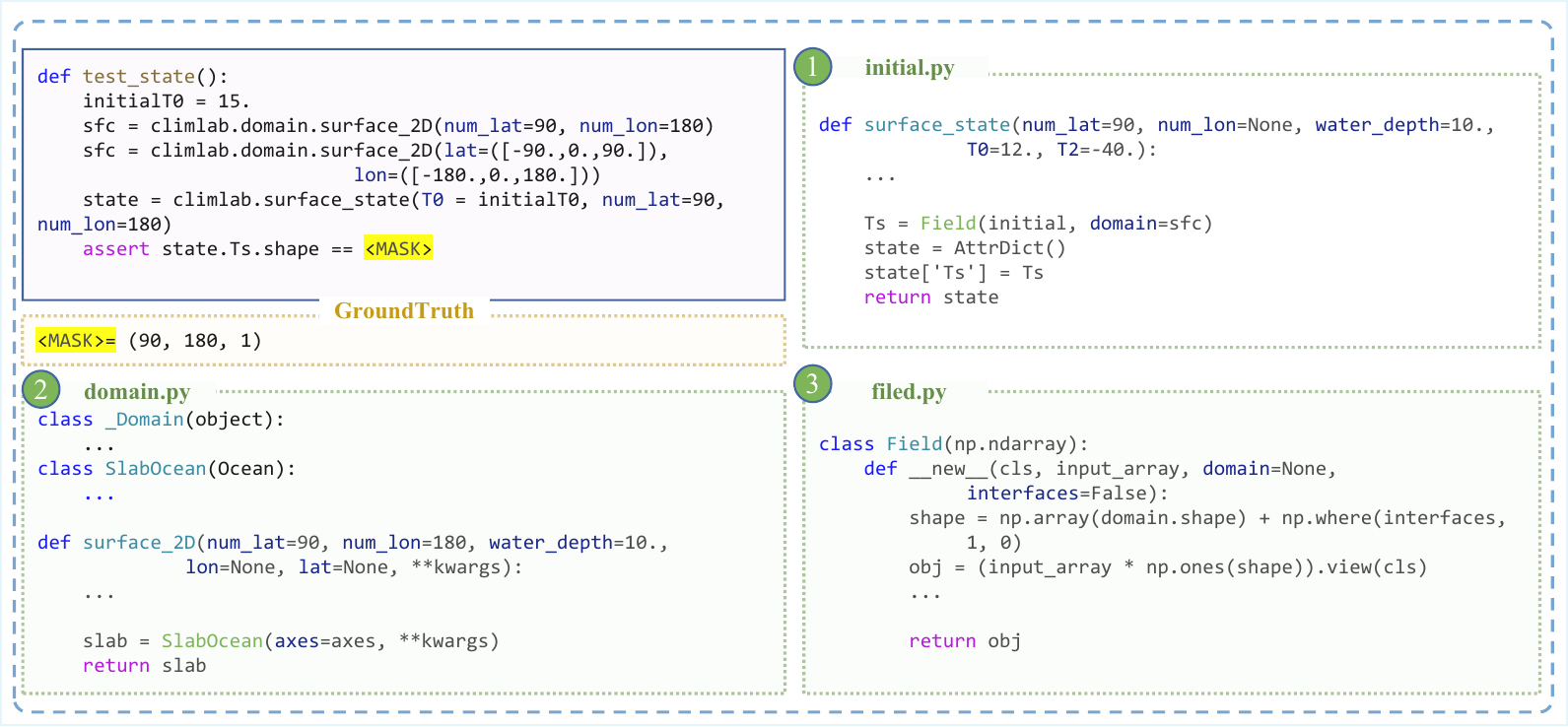}
\caption{\revision{Representative example. LLMs need to understand the code through the call chain order of files.}}
\label{fig:dependency_examples}
\end{figure}

\revision{\textbf{Cross-File Dependency Validation.} To ensure that our benchmark requires repository-level knowledge, we implemented a manual validation process. We randomly sampled 100 instances from our dataset and manually verified that they require repository-level knowledge to solve. Figure~\ref{fig:dependency_examples} illustrates representative example of positive cases. Predicting the output requires tracing different definition in the multi-hop dependency chain.}

\section{Experimental Design}
This section outlines the experimental methodology we employed to evaluate the repository-level code reasoning capabilities of LLMs. We detail the model selection, the strategies for providing long-context information, the core research questions guiding our investigation, and the specific settings used to ensure the reproducibility and validity of our results.

\subsection{Model Selection}
To provide comprehensive evaluation, we select a diverse range of LLMs across different scales, architectures, and access types. Our selection enables comparison between open-source and closed-source models while assessing the impact of specialized training on code reasoning capabilities. We evaluate seven state-of-the-art models as follows.

\begin{itemize}[left=10pt]
\item {Open-source Models: }
    \begin{itemize}[left=0pt]
        \item \textbf{Qwen-2.5-14B} serves as our base instruction-tuned baseline. \item \textbf{DeepSeek-R1-Distill-Qwen-14B} combines DeepSeek-R1's active thinking capabilities with efficient Qwen architecture.
        \item  \textbf{Qwen-2.5-Coder-14B} is specifically fine-tuned for software development tasks, allowing direct comparison with its base counterpart to isolate code-centric training benefits.
        \item  \textbf{Qwen3-235B-A22B} represents the cutting edge of large-scale open-weight models with 235 billion parameters and 22B activated parameters in its MoE architecture.
        \item  \textbf{DeepSeek-R1} provides advanced reasoning capabilities has shown exceptional performance on coding benchmarks.
    \end{itemize}
\item {Closed-source Models:}
    \begin{itemize}[left=0pt]
        \item \textbf{GPT-4.1-Mini}, a highly capable model from OpenAI that balances performance and efficiency.
        \item \textbf{Gemini-2.5-Flash-no-thinking} from Google offers high-throughput processing with a 1-million-token context window, optimized for speed over deep reasoning.
    \end{itemize}
\end{itemize}

These models allow us to investigate whether advanced repository-level reasoning emerges from model scale, specific architectures, or training methodologies across different model categories.

\subsection{Experimental Settings}
Across all experiments, we maintain a consistent set of generation parameters to ensure a fair comparison. To account for the stochastic nature of language models, we generate 5 samples for each task instance (\(n=5\)). For decoding, we use a temperature of 0.7 and a top-p value of 0.9 to encourage diversity while maintaining quality. The experiment is conducted on a server running CentOS Linux release 7.9.2009 (Core) and equipped with 1 Intel(R) Xeon(R) Platinum 8376H CPU @ 2.60GHz, and 2 NVIDIA A800 with 80GB memory.

\subsection{Evaluation Metrics}

\subsubsection{Output Prediction}
For this task, we use \textbf{Pass@k}, a standard metric in code generation that measures the percentage of problems for which at least one of the top \(k\) generated samples is correct. A generated output is considered correct if it achieves an exact match (EM) with any of the ground-truth answers. We report results for \(k=1\) and \(k=5\). Formally, given a set of problems \(\mathcal{P}\), Pass@k is defined as:
\begin{equation}
\text{Pass@k} = \frac{1}{|\mathcal{P}|} \sum_{p \in \mathcal{P}} \mathbb{I}\left( \exists i \in \{1, \dots, k\} : \text{eval}(G_{p,i}, A_p) = \text{True} \right)
\end{equation}
where \(G_{p,i}\) is the \(i\)-th generated sample for problem \(p\), \(A_p\) is the set of valid answers, \(\text{eval}(\cdot)\) is the exact match evaluation function, and \(\mathbb{I}(\cdot)\) is the indicator function.

\subsubsection{Call Chain Prediction}
To evaluate the model's ability to identify the precise set of files in a call chain, we use both F1-score and EM.
\begin{itemize}
    \item \textbf{F1-Score:} This metric provides a balanced measure of precision and recall over the set of predicted file paths. Let \(S_{\text{pred}}\) be the set of unique file paths in the predicted chain and \(S_{\text{true}}\) be the set in the ground-truth chain. The F1-score is the harmonic mean of precision and recall, calculated as:

    \arxiv{\begin{equation}
    \text{Precision} = \frac{|S_{\text{pred}} \cap S_{\text{true}}|}{|S_{\text{pred}}|},\quad
    \text{Recall} = \frac{|S_{\text{pred}} \cap S_{\text{true}}|}{|S_{\text{true}}|},\quad
    \text{F1} = 2 \cdot \frac{\text{Precision} \cdot \text{Recall}}{\text{Precision} + \text{Recall}}
    \end{equation}
    }
    
    \revision{Note that we adopt the macro-averaged computation for F1-Score. In our evaluation, each task instance is sampled multiple times ($n=5$), and for each sample, we compute independent Precision, Recall, and F1 scores. The final reported metrics are macro-averaged across all instances, ensuring each test case is weighted equally regardless of call chain length. Specifically, let $P_i$, $R_i$, and $F1_i$ denote the precision, recall, and F1 for the $i$-th instance. The reported values are:}

    \arxiv{
    \begin{equation}
    \bar{P} = \frac{1}{N}\sum_{i=1}^{N} P_i,\quad
    \bar{R} = \frac{1}{N}\sum_{i=1}^{N} R_i,\quad
    \overline{F1} = \frac{1}{N}\sum_{i=1}^{N} F1_i
    \end{equation}
    }

    \revision{Note that the macro-averaged F1 ($\overline{F1}$) is generally not equal to the F1 computed from macro-averaged precision and recall ($2\bar{P}\bar{R}/(\bar{P}+\bar{R})$). We report macro-averaged F1 to ensure equal weighting of all test cases.}
    
    \item \textbf{Exact Match (EM):} This is a stricter metric that measures the fraction of predictions where the predicted sequence of file paths is identical to the ground-truth sequence.
\end{itemize}

\subsubsection{Context Construction Strategies}
\noindent A central challenge in repository-level reasoning is providing the model with correct and sufficient context from a large corpus of files. We design distinct context acquisition strategies to systematically evaluate model performance under different conditions, tailored to each task.

\paragraph{For the Output Prediction Task}
\noindent We investigate two primary strategies representing realistic and ideal scenarios. The \textbf{Retrieval-based Context (Retrieval)} strategy simulates a practical scenario where a model must rely on an information retrieval system. For a given target test function, we treat every other source file in the repository as a document and use BM25, a standard term-based retrieval method, to score and rank these documents based on their relevance to the test function's code. The top-ranked documents are then concatenated to form the context, truncated to a predefined token limit. The \textbf{Oracle-based Context (Oracle)} strategy establishes an experimental upper bound for performance by providing the model with perfect, ground-truth context. Leveraging the call chains collected during our benchmark construction, we provide the exact source code of all files that are dynamically invoked during the execution of the target test function. This context is minimal yet complete, containing all necessary information and no extraneous noise.

\paragraph{For the Call Chain Prediction Task}
\noindent The context for this task is designed to explicitly test a model's ability to identify relevant files from a noisy background. We create the context by mixing ground-truth ``signal'' files with irrelevant ``noise'' files. The context consists of a subset of files from the ground-truth call chain (the signal) combined with randomly sampled files from elsewhere in the repository (the noise). By varying the ratio of signal-to-noise files within a fixed total context length, we can precisely measure the model's ability to discern and reconstruct the correct dependency graph. \revision{We also adopt the Oracle setting, which means that the context does not contain noise. The purpose is to evaluate the optimal performance of the model on this task.}

\subsection{Research Questions}
Our experimental design is guided by five primary research questions:
\begin{itemize}
    \item \textbf{RQ1: How do LLMs perform with sufficient reference context?}
    We evaluate LLMs on Output Prediction under oracle context (10\kmath{} tokens of ground-truth files) to isolate pure reasoning capability from retrieval noise.
    
    \item \textbf{RQ2: How do LLMs perform in understanding project structure and dependency relationships?}
    We assess LLMs' architectural comprehension via Call Chain Prediction, evaluating whether models can reconstruct dependency graphs from incomplete or noisy repository context.

    \item \textbf{RQ3: To what extent do LLMs rely on memorization instead of actual reasoning?}
    We compare model performance on original versus I/O rewritten but logically identical code to quantify the degree to which LLMs depend on memorized patterns rather than genuine reasoning.

    \item \textbf{RQ4: Does providing more context always improve repository-level reasoning?}
    We compare model performance under retrieval-based contexts of different sizes (10\kmath{} vs. 30\kmath{} tokens) to analyze the trade-off between information completeness and contextual noise.

    \item \textbf{RQ5: What are the failure modes in repository-level code reasoning?}
    We conduct a qualitative analysis of incorrect predictions to identify underlying causes of model failures beyond quantitative metrics.
\end{itemize}

\section{EVALUATION}
This section presents the empirical results of our experiments, which are designed to address a series of research questions. We investigate the performance of various LLMs on repository-level code reasoning, starting with their ideal capabilities and then examining the impact of key factors such as call chain identification, data memorization, context length, and noise. Finally, we conclude with a qualitative case study of common error types.

\subsection{RQ1. Oracle Context Performance}
We first establish a performance baseline by providing LLMs with ideal context conditions. We evaluate all LLMs on the Output Prediction task using Oracle-based Context strategy with 10\kmath{} tokens containing exactly the ground-truth files from the execution trace. This eliminates information retrieval challenges and focuses purely on reasoning capabilities.

As shown in Table 1~\ref{tab:rq1_oracle_performance}, even with perfect, exact context, the best-performing model (DeepSeek-R1) achieves only 69.1\% Pass@1 accuracy. This indicates that repository-level reasoning (distinct from retrieval) is a fundamental limitation, as nearly one-third of tasks remain unsolved despite full contextual access. Consistent with prior work, larger models (e.g., GPT-4.1-Mini, DeepSeek-R1) and code-specific variants (Qwen-2.5-Coder-14B at 51.9\% vs. Qwen-2.5-14B at 50.8\%) show modest gains, confirming that scale improves performance incrementally. However, the persistent gap to completion—despite optimal input—suggests current models lack robust mechanisms for cross-file reasoning. The 11.2\% improvement from Pass@1 to Pass@5 further implies that models generate multiple plausible paths but struggle to reliably select the correct one on the first attempt.

Reasoning-specific training alone can be counterproductive. DeepSeek-R1-Distill-Qwen-14B performs worse than its base model, indicating potential degradation in fine-grained code understanding. \revision{This phenomenon may arise from knowledge distillation exacerbating the verbosity of chain-of-thought reasoning. The model becomes overly reliant on stepwise execution trajectories, yet fails to abstract the final output structure. This behavior aligns with the elevated identifier confusion error rate observed in RQ5. By comparison, the full DeepSeek-R1 model with 671B parameters benefits from both larger model scale and dedicated reasoning training, indicating that targeted enhancements to reasoning ability rely on sufficient model capacity to achieve effectiveness.}

\begin{table}[htbp]
\centering
\caption{Performance on the Output Prediction task under ideal Oracle context (10\kmath{} tokens). This table shows the upper-bound reasoning capability of each model on the original data.}
\label{tab:rq1_oracle_performance}
\begin{tabular}{lcc}
\toprule
\textbf{Model} & \textbf{Pass@1} & \textbf{Pass@5} \\
\midrule
\textbf{Qwen-2.5-14B} & 0.508 & 0.610 \\
\textbf{R1-Distill-Qwen-14B} & 0.468 & 0.576 \\
\textbf{Qwen-2.5-Coder-14B} & 0.519 & 0.663 \\
\textbf{GPT-4.1-Mini} & 0.666 & 0.723 \\
\textbf{Gemini-2.5-Flash} & 0.624 & 0.669 \\
\textbf{Qwen3-235B} & 0.607 & 0.657 \\
\textbf{DeepSeek-R1} & 0.691 & 0.803 \\
\bottomrule
\end{tabular}
\end{table}

\begin{center}
\begin{myboxc}{\textbf{RQ1 Summary: }
The results of DeepSeek-R1 demonstrate that LLMs face a performance ceiling in repository-level reasoning, even under ideal conditions with perfect context. While model scale and code-specific training yield marginal improvements, they are insufficient to close this gap, pointing to a core need for architectures that explicitly support structured, cross-file inference.
}
\end{myboxc}
\end{center}

\subsection{RQ2. Dependency Relationship Understanding}

\begin{table}[htbp]
\centering
\caption{Performance on the Call Chain Prediction task with varying signal-to-noise ratios in a 20\kmath{} token context. \revision{The Oracle setting provides ground-truth files only, establishing an upper bound.} The results highlight a common pattern of high precision but low recall.}
\label{tab:rq2_call_chain}
\begin{tabular}{lccccc}
\toprule
\textbf{Model} & \textbf{Signal Ratio} & \textbf{Precision} & \textbf{Recall} & \textbf{F1} & \textbf{EM} \\
\midrule
\multirow{3}{*}{\textbf{Qwen-2.5-14B}} & 25\% & 0.195 & 0.182 & 0.167 & 0.047 \\
& 50\% & 0.233 & 0.215 & 0.189 & 0.036 \\
& \revision{Oracle} & \revision{0.413} & \revision{0.292} & \revision{0.342} & \revision{0.140} \\
\midrule
\multirow{3}{*}{\textbf{R1-Distill-Qwen-14B}} & 25\% & 0.786 & 0.346 & 0.417 & 0.154 \\
& 50\% & 0.669 & 0.321 & 0.377 & 0.155 \\
& \revision{Oracle} & \revision{0.908} & \revision{0.370} & \revision{0.526} & \revision{0.176} \\
\midrule
\multirow{3}{*}{\textbf{Qwen-2.5-Coder-14B}} & 25\% & 0.195 & 0.151 & 0.158 & 0.065 \\
& 50\% & 0.241 & 0.175 & 0.178 & 0.065 \\
& \revision{Oracle} & \revision{0.295} & \revision{0.196} & \revision{0.235} & \revision{0.102} \\
\midrule
\multirow{3}{*}{\textbf{GPT-4.1-Mini}} & 25\% & 0.801 & 0.312 & 0.388 & 0.136 \\
& 50\% & 0.814 & 0.318 & 0.397 & 0.133 \\
& \revision{Oracle} & \revision{0.912} & \revision{0.355} & \revision{0.511} & \revision{0.160} \\
\midrule
\multirow{3}{*}{\textbf{Gemini-2.5-Flash}} & 25\% & 0.649 & 0.433 & 0.465 & 0.176 \\
& 50\% & 0.674 & 0.475 & 0.505 & 0.186 \\
& \revision{Oracle} & \revision{0.817} & \revision{0.558} & \revision{0.663} & \revision{0.213} \\
\midrule
\multirow{3}{*}{\textbf{Qwen3-235B}} & 25\% & 0.820 & 0.361 & 0.441 & 0.148 \\
& 50\% & 0.836 & 0.372 & 0.457 & 0.148 \\
& \revision{Oracle} & \revision{0.913} & \revision{0.387} & \revision{0.543} & \revision{0.179} \\
\midrule
\multirow{3}{*}{\textbf{DeepSeek-R1}} & 25\% & 0.785 & 0.339 & 0.408 & 0.181 \\
& 50\% & 0.821 & 0.373 & 0.447 & 0.186 \\
& \revision{Oracle} & \revision{0.952} & \revision{0.357} & \revision{0.520} & \revision{0.167} \\
\bottomrule
\end{tabular}
\end{table}

\begin{figure}[ht!]
    \centering
    \includegraphics[width=\textwidth]{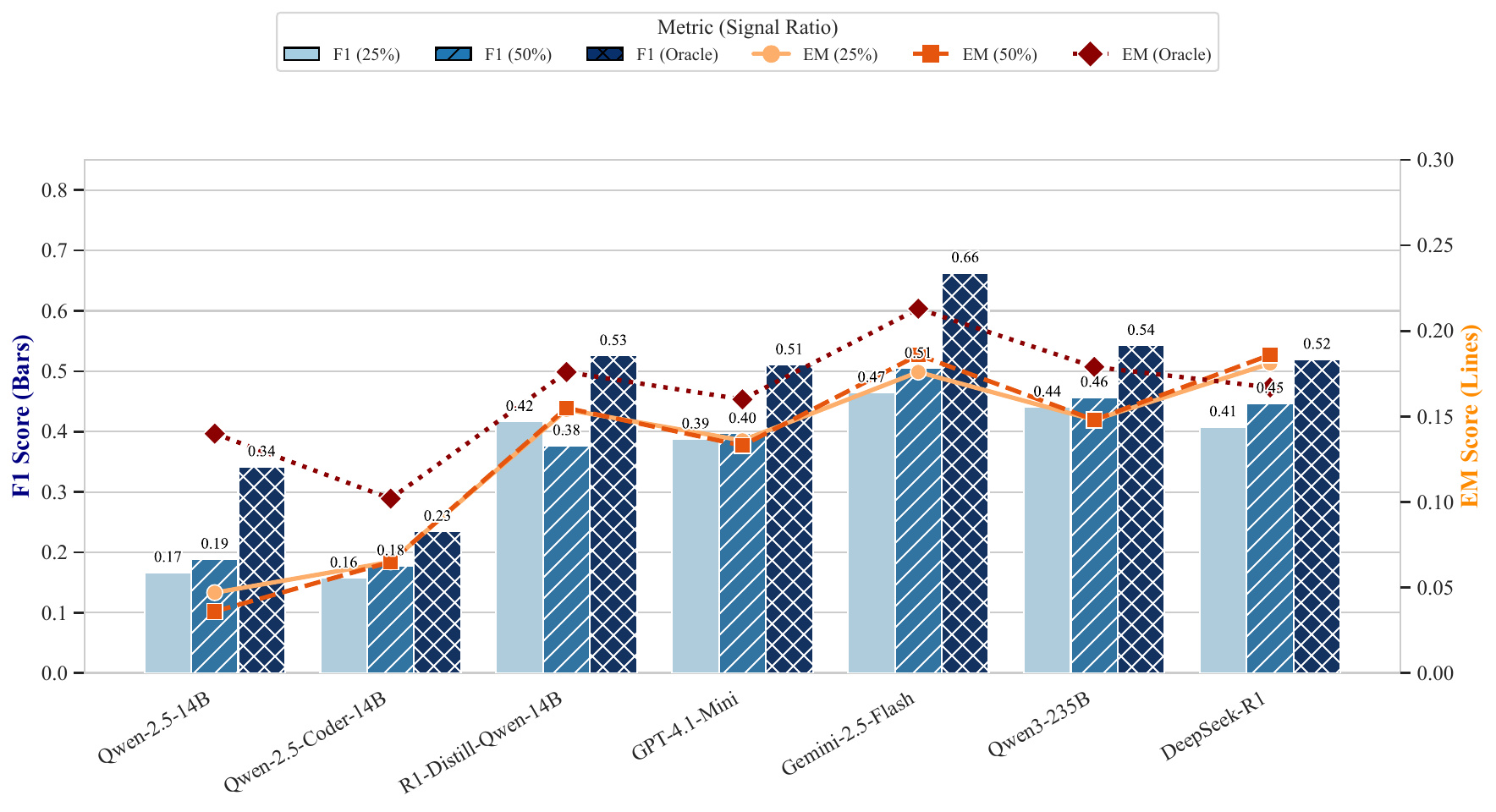}
    \caption{Comparison of F1 Scores (Bar Chart, Left Y-axis) and EM Scores (Line Chart, Right Y-axis) Across Models for Call Chain Prediction.}
    \label{fig:rq2}
\end{figure}

We evaluate LLMs' ability to understand repository architecture through the Call Chain Prediction task. LLMs receive a noisy file pool (20\kmath{} tokens) and must identify the correct sequence of files invoked by a test function. This measures their capability to trace control flow and reconstruct dependency relationships.

Table~\ref{tab:rq2_call_chain} reveals that reconstructing multi-file dependency graphs is extremely challenging for all LLMs. A consistent pattern emerges: high precision but low recall. Larger LLMs like GPT-4.1-Mini and Qwen3-235B achieve precision above 80\% but recall below 40\%. This indicates they can identify some direct dependencies (likely through import statements) but struggle with deep, multi-hop traversal required for complete call chain reconstruction.

The low F1 scores and sub-20\% exact match (EM) scores demonstrate that no LLM can consistently identify complete call chains. Even the best-performing LLM (DeepSeek-R1 in 25\% \& 50\% signal ratio) achieves only 18.6\% EM score. Figure~\ref{fig:rq2} visualizes the relationship between F1 and EM scores across LLMs, highlighting the substantial gap between partial and complete understanding.

\revision{As shown in the oracle setting, even with perfect context, the best-performing model (Gemini-2.5-Flash) achieves only 66.3\% F1 and 21.3\% EM. This reveals that the challenge in Call Chain Prediction stems not only from noisy context but also from the fundamental difficulty of tracing multi-hop dependencies. The Oracle results show improved precision (most models exceed 80\%) but recall remains limited, indicating that models struggle to identify the complete set of invoked files even when no distractors are present.}

This suggests a fundamental difference between micro-level code execution and macro-level structure comprehension. Current LLMs excel at local reasoning but lack the architectural understanding needed for repository-level tasks.

\begin{center}
\begin{myboxc}{\textbf{RQ2 Summary: }
LLMs show limited ability to understand project structure and dependencies. They can identify some direct relationships with high precision but fail to perform complete multi-hop traversal, achieving low recall and poor exact match scores. \revision{Even under Oracle conditions with perfect context, models struggle to achieve high recall, confirming that multi-hop dependency tracing is a bottleneck.
}}
\end{myboxc}
\end{center}

\subsection{RQ3. Memorization vs. Reasoning}

\begin{table}[t]
\centering
\caption{Comparison of Pass@1 and Pass@5 scores on Original vs. I/O-Rewritten data in the 10\kmath{} Oracle context.}
\label{tab:rq3_memorization}
\begin{tabular}{lcccc}
\toprule
\textbf{Model} & \multicolumn{2}{c}{\textbf{Original}} & \multicolumn{2}{c}{\textbf{I/O-Rewritten}} \\
\cmidrule(lr){2-3} \cmidrule(lr){4-5}
 & \textbf{Pass@1} & \textbf{Pass@5} & \textbf{Pass@1} & \textbf{Pass@5} \\
\midrule
\textbf{Qwen-2.5-14B} & 0.508 & 0.610 & 0.420 & 0.529 \\
\textbf{R1-Distill-Qwen-14B} & 0.468 & 0.576 & 0.438 & 0.535 \\
\textbf{Qwen-2.5-Coder-14B} & 0.519 & 0.663 & 0.459 & 0.603 \\
\textbf{GPT-4.1-Mini} & 0.666 & 0.723 & 0.588 & 0.656 \\
\textbf{Gemini-2.5-Flash} & 0.624 & 0.669 & 0.584 & 0.645 \\
\textbf{Qwen3-235B} & 0.607 & 0.657 & 0.563 & 0.619 \\
\textbf{DeepSeek-R1} & 0.691 & 0.803 & 0.650 & 0.766 \\
\bottomrule
\end{tabular}
\end{table}

\begin{figure}[ht!]
    \centering
    \includegraphics[width=\textwidth]{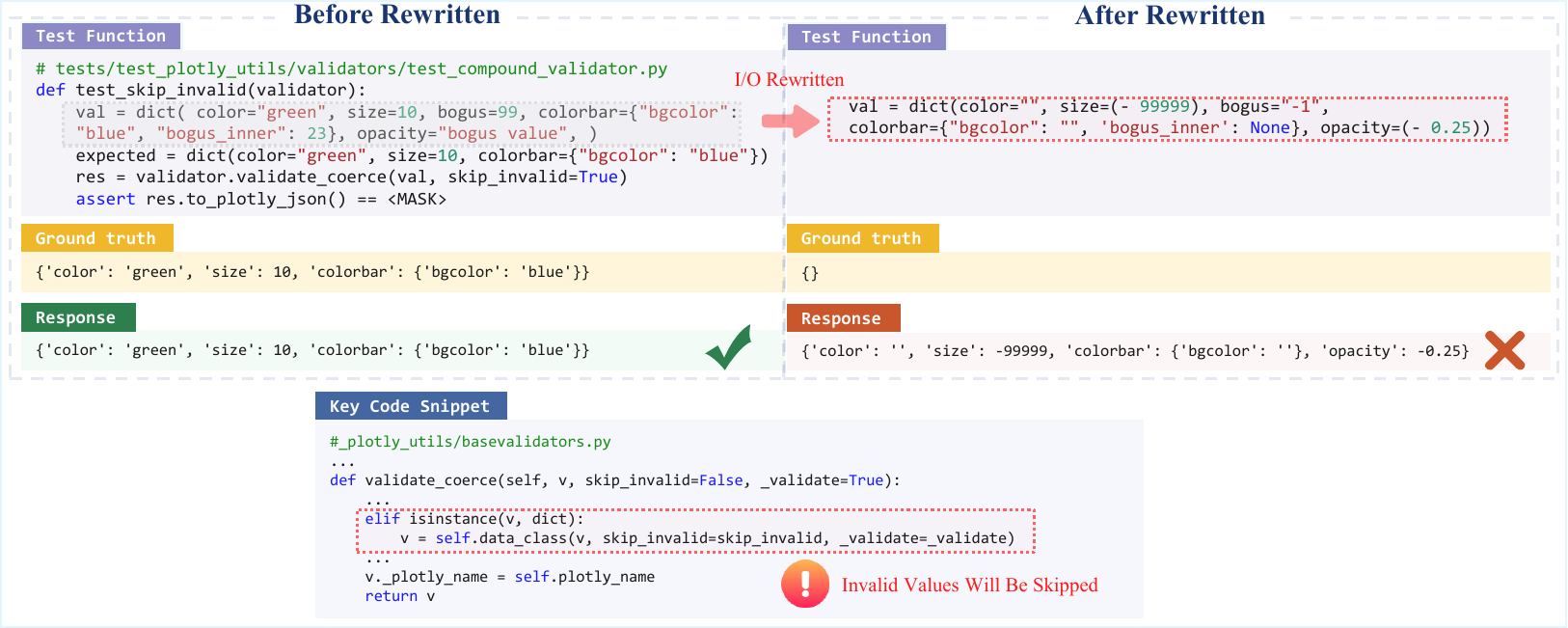}
    \caption{Case of reasoning influenced by memorization.}
    \label{fig:rq3case}
\end{figure}

To distinguish genuine reasoning from pattern matching, we compare LLM performance on original versus rewritten datasets. The ``I/O-Rewritten'' version is algorithmically generated to be logically identical but syntactically different (different Input/Output). We evaluate both versions using the Output Prediction task under 10\kmath{} Oracle context.

Table~\ref{tab:rq3_memorization} shows universal performance drops on rewritten code, confirming that memorization contributes to LLM success. GPT-4.1-Mini drops from 66.6\% to 58.8\% Pass@1, while DeepSeek-R1 decreases from 69.1\% to 65.0\%. This indicates LLMs rely on ``cognitive shortcuts'' from training data rather than pure logical deduction.

The performance gaps suggest LLMs handle familiar code patterns well but become brittle with \revision{unfamiliar code patterns}. For example, they might correctly process a frequently-seen function call but fail when the same logic appears in different control flows or variable naming schemes.

Figure~\ref{fig:rq3case} illustrates this brittleness. On original data, after dict initialization, LLMs correctly understand the \texttt{validate\_coerce} function validates dictionary properties. However, when the rewritten version contains invalid attributes, LLMs struggle to detect the anomaly, showing their dependence on familiar patterns rather than robust reasoning.

\begin{center}
\begin{myboxc}{\textbf{RQ3 Summary: }
All LLMs rely partially on memorization rather than pure reasoning. Universal performance drops on syntactically altered but logically identical code reveal brittleness in handling unfamiliar patterns, indicating dependence on training data shortcuts rather than robust deduction.
}
\end{myboxc}
\end{center}

\subsection{RQ4. Context Length Effects}
\begin{table}[t]
\centering
\caption{Performance on the Output Prediction task comparing 10\kmath{} and 30\kmath{} Retrieval contexts.}
\label{tab:rq4_context_volume}
\begin{tabular}{llcccc}
\toprule
\textbf{Model} & \textbf{Data Type} & \multicolumn{2}{c}{\textbf{10\kmath{} Retrieval}} & \multicolumn{2}{c}{\textbf{30\kmath{} Retrieval}} \\
\cmidrule(lr){3-4} \cmidrule(lr){5-6}
 & & \textbf{Pass@1} & \textbf{Pass@5} & \textbf{Pass@1} & \textbf{Pass@5} \\
\midrule
\multirow{2}{*}{\textbf{Qwen-2.5-14B}} & Original & 0.360 & 0.467 & 0.403 & 0.506 \\
 & I/O-Rewritten & 0.347 & 0.458 & 0.384 & 0.498 \\
\midrule
\multirow{2}{*}{\textbf{R1-Distill-Qwen-14B}} & Original & 0.341 & 0.470 & 0.367 & 0.494 \\
 & I/O-Rewritten & 0.346 & 0.439 & 0.360 & 0.473 \\
\midrule
\multirow{2}{*}{\textbf{Qwen-2.5-Coder-14B}} & Original & 0.404 & 0.540 & 0.378 & 0.521 \\
 & I/O-Rewritten & 0.396 & 0.526 & 0.363 & 0.498 \\
\midrule
\multirow{2}{*}{\textbf{GPT-4.1-Mini}} & Original & 0.541 & 0.621 & 0.550 & 0.631 \\
 & I/O-Rewritten & 0.516 & 0.607 & 0.532 & 0.629 \\
\midrule
\multirow{2}{*}{\textbf{Gemini-2.5-Flash}} & Original & 0.539 & 0.587 & 0.520 & 0.579 \\
 & I/O-Rewritten & 0.519 & 0.571 & 0.515 & 0.586 \\
\midrule
\multirow{2}{*}{\textbf{Qwen3-235B}} & Original & 0.507 & 0.569 & 0.519 & 0.580 \\
 & I/O-Rewritten & 0.499 & 0.556 & 0.512 & 0.577 \\
\midrule
\multirow{2}{*}{\textbf{DeepSeek-R1}} & Original & 0.585 & 0.704 & 0.583 & 0.703 \\
 & I/O-Rewritten & 0.572 & 0.691 & 0.568 & 0.690 \\
\bottomrule
\end{tabular}
\end{table}

We investigate whether more context consistently improves performance by comparing 10\kmath{} versus 30\kmath{} token contexts. Both use a Retrieval-based Context strategy on the Output Prediction task, representing realistic scenarios where context may contain irrelevant information.

Table~\ref{tab:rq4_context_volume} reveals that increased context length produces mixed effects depending on LLM architecture. Larger, more capable LLMs like GPT-4.1-Mini and Qwen3-235B show slight but consistent improvements with 30\kmath{} context, suggesting effective noise filtering capabilities. However, even powerful LLMs like DeepSeek-R1 experience slight degradation, indicating that excessive context can dilute attention mechanisms.

Smaller LLMs show more varied responses. Base Qwen-2.5-14B improves notably (36.0\% to 40.3\% Pass@1), likely because longer context increases the probability of retrieving crucial code snippets. Conversely, Qwen-2.5-Coder-14B shows decreased performance (40.4\% to 37.8\% Pass@1), suggesting that specialized code training makes LLMs more sensitive to irrelevant noise.

These results highlight the importance of robust attention mechanisms. While more context can provide additional relevant information, it also introduces noise that can overwhelm LLMs without strong filtering capabilities.

\begin{center}
\begin{myboxc}{\textbf{RQ4 Summary: }
More context does not always improve performance. While a longer context may contain more relevant information, it also introduces noise. Larger LLMs generally handle this trade-off better, while smaller and specialized LLMs show varied sensitivity to context length increases.
}
\end{myboxc}
\end{center}

\subsection{RQ5. Error Analysis}

\revision{To understand the underlying causes of model failures beyond quantitative metrics, we conduct an error analysis on both Output Prediction and Call Chain Prediction tasks. We manually examine 200 failed samples from Output Prediction and 100 from Call Chain Prediction, categorizing errors by their root causes. This qualitative investigation reveals failure modes that quantitative scores alone cannot capture.}

\subsubsection{Output Prediction Error Analysis}

\begin{figure}[ht!]
\centering
\includegraphics[width=\textwidth]{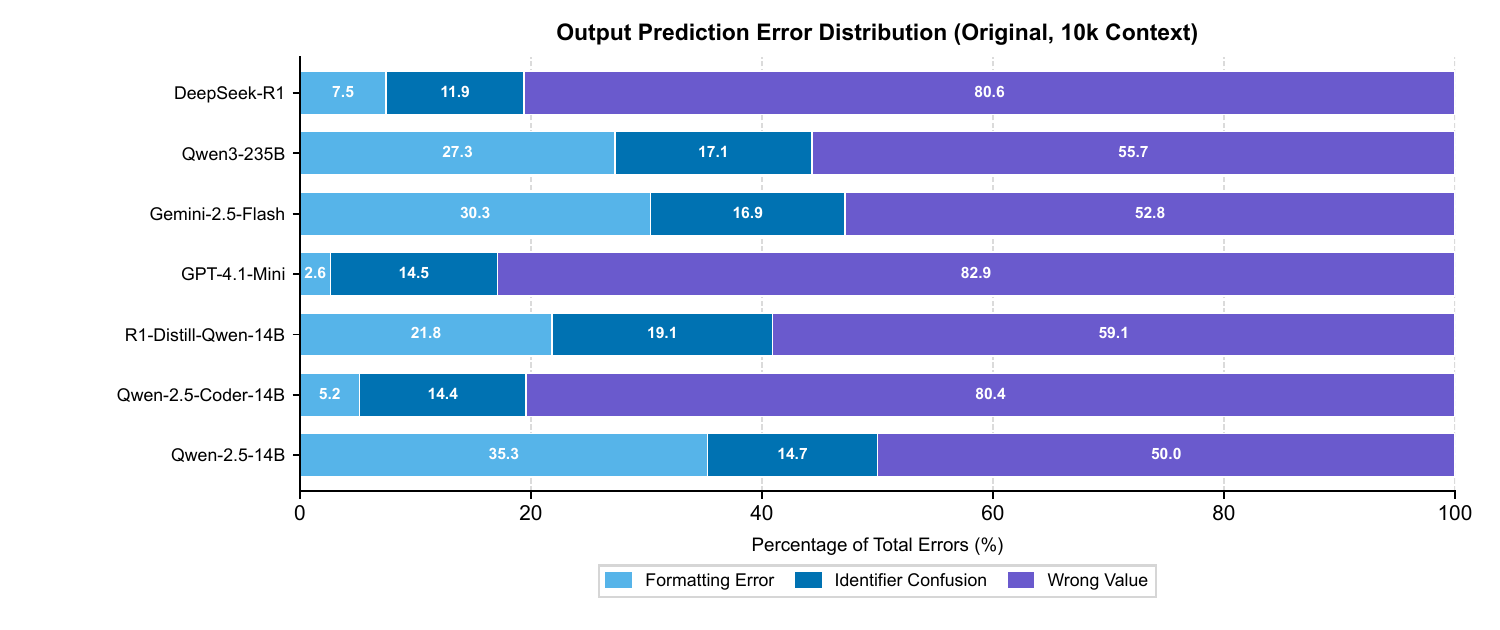}
\caption{\revision{Distribution of error types in Output Prediction task. The three categories are: Formatting errors, Identifier Confusion, and Wrong Value.}\protect\footnotemark}
\label{fig:op_error_distribution}
\end{figure}

\revision{We classify Output Prediction errors into three categories. Figure~\ref{fig:op_error_distribution} presents the error distribution across all evaluated models under 10k context with original data.}

\newcounter{fnspace}
\setcounter{fnspace}{\value{footnote}}

\afterpage{\footnotetext{\arxiv{Due to space limitations, only figures under one experimental setting are presented. We provide some analysis under other
experimental settings in the paper, and other figures are available in the online package.}}}

\begin{itemize}

\item \textbf{Formatting or Extraction Errors.}
This category covers (1)~\textit{instruction-following failures} (extraneous text, code 
blocks, or unintended newlines) and (2)~\textit{format mismatches} where the semantic value 
is correct but string representation differs. Code-specialized models substantially 
reduce this error (Qwen-2.5-Coder-14B: 5.2\% vs.\ Qwen-2.5-14B: 35.3\%), demonstrating 
that domain-specific training improves output format compliance. As shown in 
Figure~\ref{fig:rq5case1}, a model may correctly identify the return value as ``abc'' yet 
output the bare identifier rather than the string literal \texttt{`abc'}.

\begin{figure}[ht!]
\centering
\includegraphics[width=1.0\textwidth]{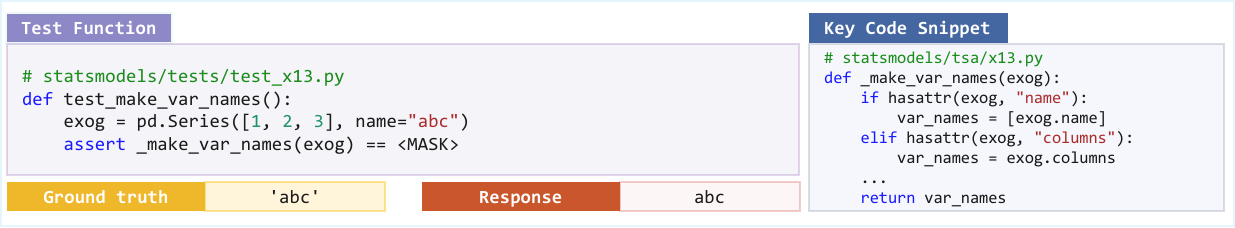}
\caption{Example of Formatting/Extraction Error: The model correctly understands that 
the function returns the name ``abc'' but fails to format it as a string literal 
\texttt{`abc'}, instead outputting the bare identifier \texttt{abc}.}
\label{fig:rq5case1}
\end{figure}

\item \textbf{Identifier Confusion.}
This error occurs when the model predicts a symbolic expression (variable name, attribute 
chain, or indexed access) instead of the concrete literal ground truth, indicating 
confusion between a reference and its runtime value. As Figure~\ref{fig:rq5case2} 
illustrates, a model may correctly identify an internal sub-component value (e.g., 
geometry type \texttt{`Point'}) yet fail to reconstruct the enclosing dictionary the 
function actually returns. Notably, reasoning-focused models (R1-Distill-Qwen-14B: 21.8\%) 
show higher rates than code-specialized ones (Qwen-2.5-Coder-14B: 5.2\%), suggesting that 
explicit chain-of-thought reasoning can over-commit to concrete execution traces at the 
expense of recognizing the final return structure.

\begin{figure}[t]
\centering
\includegraphics[width=\textwidth]{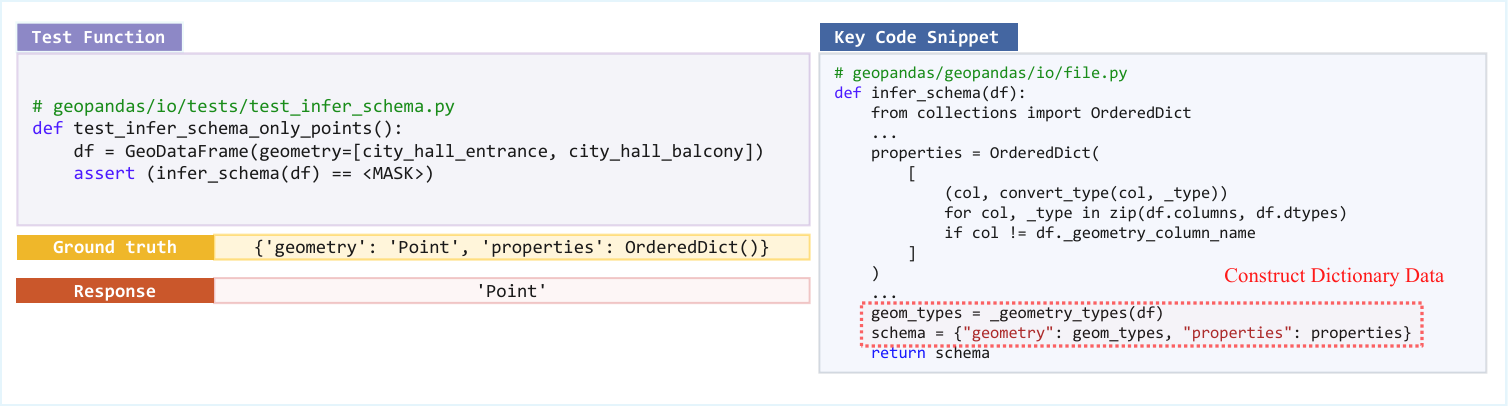}
\vspace{-10pt}
\caption{Example of Identifier Confusion: The model correctly identifies the value of a 
specific internal variable (e.g., the geometry type 'Point') but fails to recognize that 
the function's final output should be a reconstructed dictionary structure, confusing a 
sub-component's value with the overall return object.}
\label{fig:rq5case2}
\end{figure}

\item \textbf{Wrong Value.}
This dominant category (50.0--82.9\% across all models) captures cases where the model 
produces correctly formatted output but with incorrect content due to flawed reasoning. 
Figure~\ref{fig:rq5_undefined_var} exemplifies a control-flow failure: the model predicts 
\texttt{expected\_tuples}, a variable defined \emph{after} the assertion, revealing an 
inability to respect program execution order. Broader underlying causes include weak 
cross-file state tracking, misresolution of object-oriented dispatch, and 
library-specific semantic misunderstandings (e.g., pytest fixtures, NumPy broadcasting).

\begin{figure}[t]
\centering
\includegraphics[width=0.7\textwidth]{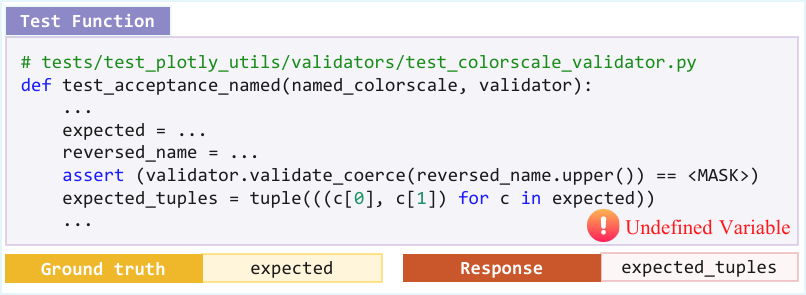}
\vspace{-10pt}
\caption{Example of Wrong Value Error: The model predicts \texttt{expected\_tuples}, a 
variable that is defined after the assertion statement, demonstrating failure to respect 
program execution order and control flow.}
\vspace{-10pt}
\label{fig:rq5_undefined_var}
\end{figure}

\end{itemize}

\subsubsection{Call Chain Prediction Error Analysis}

\begin{figure}[ht!]
    \centering
    \includegraphics[width=\linewidth]{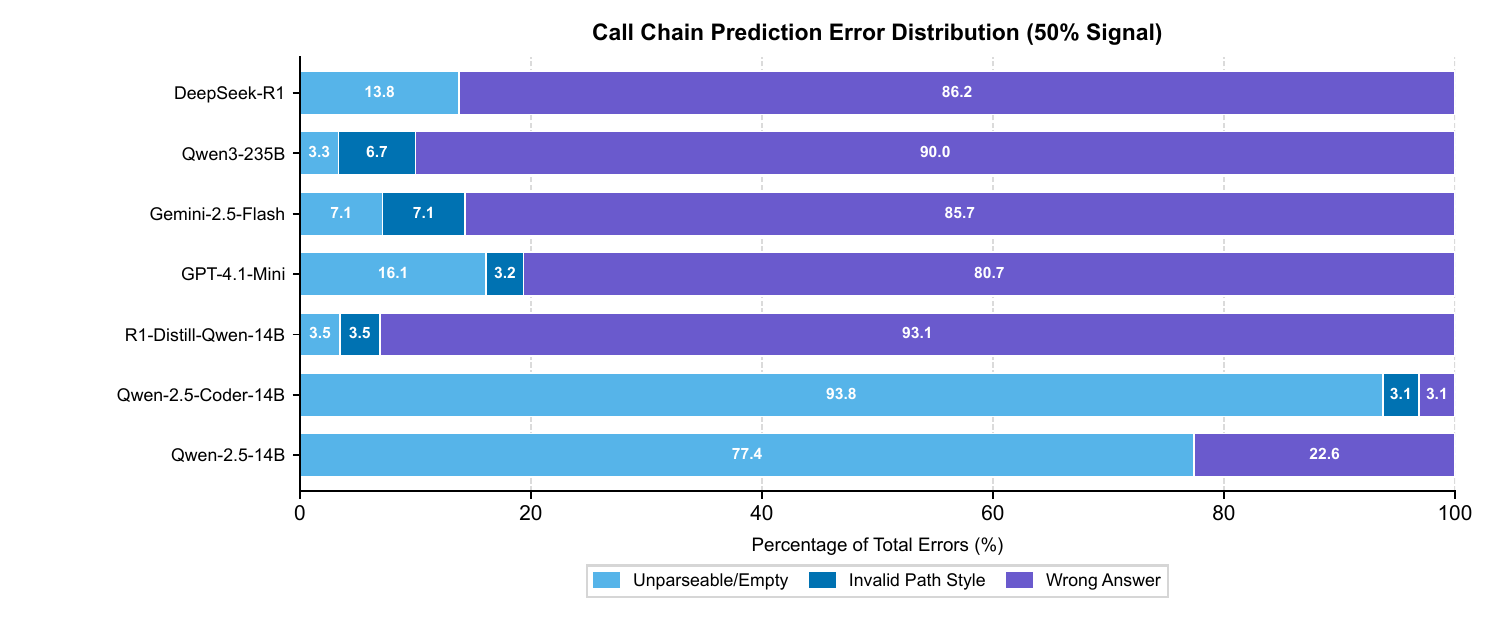}
    \caption{\revision{Distribution of error types in Call Chain Prediction task. The three categories are: Unparseable/Empty, Invalid Path Style, and Wrong Answer.}\protect\hyperlink{Hfootnote.\arabic{fnspace}}{\textsuperscript{\arabic{fnspace}}}}
    \label{fig:ccp_error_distribution}
\end{figure}

\revision{We classify Call Chain Prediction errors into three categories based on our analysis of 100 failed samples. Figure~\ref{fig:ccp_error_distribution} shows the distribution of error types across models under 50\% signal ratios.}

\revision{Three patterns emerge from the error distribution. 
\textbf{Unparseable/Empty} errors concentrate in smaller models under noisy context 
(Qwen-2.5-14B: 77.4\%; Qwen-2.5-Coder-14B: 93.8\% at 50\% signal), while larger 
models maintain structured outputs (3.3--16.1\%), indicating that model scale governs 
robustness to noise in instruction-following.
\textbf{Invalid Path Style} errors remain consistently low ($\leq$8\%) across all models 
and settings, confirming that path-formatting conventions are reliably inferred from 
the provided context.
\textbf{Wrong Answer} errors dominate for larger models (80.7--93.1\%), and—critically—
persist even under Oracle context where all provided files are relevant (73.3--96.5\%). 
This suggests that the core bottleneck is architectural reasoning rather than noise 
filtering: models identify directly imported files but fail to trace deeper multi-hop 
dependencies even with perfect context.}

\begin{center}
\begin{myboxc}{\textbf{RQ5 Summary: }
\revision{Our error analysis reveals consistent failure patterns across both tasks. For Output Prediction, Wrong Value errors dominate, showing that core execution reasoning is a key bottleneck even when formatting improves. For Call Chain Prediction, noise mainly harms smaller models. These results indicate that cross-file state tracking and architectural reasoning remain limitations in repository-level code understanding.}}
\end{myboxc}
\end{center}

\section{Threats to Validity}
We identify the following key threats to the validity of our study.

\subsection{\textbf{Internal Validity}} 
\ \ \ \ \textbf{Data Leakage:} A primary threat is that models may have encountered similar code logic from repositories during pre-training. This could lead them to predict the output via memorization rather than genuine reasoning. To mitigate this, we implemented an I/O rewriting pipeline that generates ``new data'' for evaluation. From the evaluation, it seems that I/O rewriting is insufficient. In the future, we can try refactoring functions and refactoring call chains. 

\revision{\textbf{I/O Rewriting Limitations:} Our I/O rewriting only modifies input values and expected outputs, preserving all other code logic to ensure logical equivalence and transformation reliability. More sophisticated refactoring, such as function restructuring or call chain reorganization could more effectively challenge memorization; we leave such extensions to future work.}

\textbf{Benchmark Correctness:} The reliability of our benchmark hinges on our automated data construction pipeline, including the dynamic tracer and AST parser. Potential bugs in these components could introduce errors into the ground-truth data. To address this, we rigorously tested these tools across diverse codebases and implemented multiple validation checks throughout the pipeline to ensure the accuracy of call chains and extracted data. \revision{To ensure the accuracy of the data as much as possible, we plan to conduct more thorough validation in future iterations of the benchmark.}

\textbf{Evaluation Metrics:} We evaluate via static matching rather than execution to avoid prohibitive overhead from repeated sampling. This may have an impact on the validity of the experimental results. But we have tried to collect and supplement ground-truth in the data collection stage, and have made efforts to reduce this impact.

\subsection{\textbf{External Validity}} 
\ \ \ \ \textbf{Generalizability:} Our benchmark is limited to Python, whose architectural patterns may not generalize to other languages.
To mitigate this, we select projects spanning diverse domains and scales, enhancing internal variation. Future work will extend the evaluation to multiple languages. 

\textbf{Retrieval Method:} Our study employs BM25 as the sole retrieval method for creating realistic contexts. More advanced techniques, such as vector-based retrieval, might yield different performance results. Our choice of BM25, however, provides a well-established and reproducible baseline, allowing for a clear comparison of models' core reasoning abilities without confounding variables from a complex retrieval system. We plan to incorporate and compare multiple retrieval strategies in future iterations. 

\textbf{Model Selection:} The rapid evolution of LLMs means our selection, while diverse, may not capture the capabilities of the absolute latest or largest models. To mitigate this, we selected a representative suite of models spanning open-source and closed-source, varying in scale, and including both general-purpose and code-specific architectures. This ensures our findings are robust across different model types. We plan to continuously update our evaluation with new and more powerful models as they become accessible.

\section{Conclusion}

In this paper, we address the limitation of existing evaluations that focus on function-level reasoning and overlook repository-scale challenges. We introduce \bench for repository-level code reasoning and conduct a evaluation on diverse LLMs. Our results show that even under Oracle settings, models struggle with cross-file reasoning and architectural dependency understanding, indicating that scaling context windows alone is insufficient. \revision{These findings suggest the need for structured cross-file reasoning mechanisms, improved repository-level dependency modeling, and more diversified training strategies to reduce memorization and improve robustness to unfamiliar patterns.} \revision{Our benchmark construction framework is automated and scalable through environment setup and dynamic tracing, and can be extended to larger repositories and additional programming languages.} We believe \bench provides a realistic evaluation framework that can support the development of practical repository-level software engineering assistants.

\section{Data Availability}
\label{sec:open-source}
\arxiv{Our code and data are available at \url{https://github.com/DeepSoftwareAnalytics/RepoReasoner}.}